\documentclass[a4paper,UKenglish,cleveref, autoref, thm-restate]{lipics-v2021}

\pdfoutput=1 %
\hideLIPIcs %

\usepackage{booktabs}
\usepackage{numprint}
\npthousandsep{,}
\usepackage{xcolor}
\usepackage{listings}
\lstdefinelanguage{Lean4}{
  morekeywords={import,noncomputable,section,open,axiom,theorem,by,have,False,Type},
  sensitive=true,
  morecomment=[l]{--},
  morecomment=[s]{/-}{-/},
  morestring=[b]",
}
\title{LLM2SMT: Building an SMT Solver with\\Zero Human-Written Code}
\author{Mikol{\'{a}}\v{s} Janota}{Czech Technical University, Czechia \and \url{https://people.ciirc.cvut.cz/~janotmik/} }{mikolas.janota@gmail.com}{https://orcid.org/0000-0003-3487-784X}{%
The research was supported by the MEYS
within the dedicated program ERC~CZ under the project \textsf{POSTMAN} no.~LL1902,
and by the European Union under the project \textsf{ROBOPROX}
(reg.~no.~CZ.02.01.01/00/22\_008/0004590).
This article is part of the \textsf{RICAIP} project that has received funding from
the European Union's Horizon~2020 research and innovation programme under grant
agreement No~857306.
}
\author{Mirek Ol\v{s}\'{a}k}{Czech Technical University, Czechia \and \url{https://www.olsak.net/} }{mirek@olsak.net}{https://orcid.org/0000-0002-9361-1921}{%
  The author acknowledges the support of National Recovery Plan funded project MPO 60273/24/21300/21000 CEDMO 2.0 NPO.}

\authorrunning{M. Janota and M. Ol\v{s}\'{a}k} %

\Copyright{Mikol{\'{a}}\v{s} Janota and Mirek Ol\v{s}\'{a}k} %



\category{} %

\relatedversion{} %

\nolinenumbers %

\EventEditors{John Q. Open and Joan R. Access}
\EventNoEds{2}
\EventLongTitle{42nd Conference on Very Important Topics (CVIT 2016)}
\EventShortTitle{CVIT 2016}
\EventAcronym{CVIT}
\EventYear{2016}
\EventDate{December 24--27, 2016}
\EventLocation{Little Whinging, United Kingdom}
\EventLogo{}
\SeriesVolume{42}
\ArticleNo{23}

\begin{document}

\maketitle

\begin{abstract}
  Whether LLMs can reason or write software is widely debated, but whether they
  can write software that \emph{itself reasons} is largely unexplored. We
  present a case study in which an LLM coding agent builds a complete
  DPLL(T)-style SMT solver for QF\_UF with zero human-written code. The solver
  implements the Nieuwenhuis-Oliveras congruence closure algorithm, includes
  preprocessing, and emits Lean proofs for unsatisfiable instances. We describe
  the development process and key challenges, and show that the resulting solver
  is competitive on SMT-LIB benchmarks. While full proof emission for
  unsatisfiable instances appears to be beyond the capabilities of the LLM, the
  solver achieves competitive performance and passes correctness tests.
\end{abstract}

\section{Introduction}\label{sec:introduction} %

LLM-powered coding agents are now so widely used that some believe software
engineers in their traditional role will disappear~\cite{finalroundai}.
LLMs---directly or via coding agents---have also been used for
auto-formalization and for developing novel
proofs~\cite{numina,minif2fdafny,szegedy,gemini}. The question we ask is:
\textit{Can LLMs develop automated reasoning tools?}

The appeal is clear: rapid feature integration, experimentation with new
techniques, and importing methods from recent publications.
Skepticism is warranted, however---small mistakes can propagate into subtle,
hard-to-detect errors. Correctness is paramount: do LLMs understand the
underlying logic well enough to write reliable SMT solvers?

We develop a basic DPLL(T)-style~\cite{dpll_t} SMT solver with zero code
written by a human, focusing on the theory of uninterpreted functions without
quantifiers (QF\_UF)---an important theory underlying most functionality in
DPLL(T).

The solver uses CaDiCaL~\cite{cadical} via the IPASIR-UP
interface~\cite{ipasirup}, emits proofs in Lean~\cite{lean} (under some
limitations), implements the Nieuwenhuis--Oliveras congruence closure
algorithm~\cite{closure}, and integrates several preprocessing techniques. The
code is available as a GitHub
repository\footnote{\url{https://github.com/MikolasJanota/llm2smt}}.

\section{Background}\label{sec:background} %
We assume that the reader is familiar with classical first-order logic with
equality~\cite{smullyan}.

Satisfiability Modulo Theories (SMT) solving is a well-established research area
encompassing a wide range of logical theories, applications, and algorithmic
techniques. Most SMT solvers reduce the input formula to propositional
satisfiability (SAT), either \emph{eagerly}~\cite{pnueli-cav99}, where theory
reasoning is encoded directly into a propositional (Boolean) problem, or
\emph{lazily}~\cite{demoura-cade02}, where Boolean reasoning is interleaved with
theory-specific checks. These checks result in \emph{theory lemmas},
which strengthen the propositional problem.

Most modern solvers are based on the \emph{DPLL\,(T)
framework}~\cite{dpll_t,krstic-frocos07,sebastiani-jsat07}. In this
architecture, theory atoms are abstracted as Boolean variables and handled using
conventional SAT-solving techniques~\cite{marques-silva-99}. Candidate
assignments produced by the SAT solver are subsequently validated by dedicated
\emph{theory solvers}, which are required only to check conjunctions of theory
literals. If a conjunction is inconsistent in the underlying theory, the theory
solver provides conflict information to refine the Boolean search. The process
iterates until the SAT solver determines the formula to be unsatisfiable or a
theory-consistent assignment is obtained. During the search, the solver might
employ \emph{theory propagation}, forcing certain literals to specific values
under the current partial assignment.

This paper focuses on the theory of equality with uninterpreted functions
(QF-EUF), which is the basis of equality
reasoning; in particular, it is instrumental in the \emph{Nelson-Oppen
closure}~\cite{nelson-oppen-79,moura-entcs08}, which enables cooperation between
separate theory solvers under appropriate assumptions.
In the context of SMT, quantifiers are handled by \emph{quantifier
instantiation}, cf.~\cite{simplify,moskal-smt09}. Therefore, solving problems
without quantifiers (ground problems) is necessary for solving quantified
problems.

\emph{Lean}~\cite{lean} is a popular modern interactive theorem prover (ITP)
and programming language for formalizing mathematics and verifying software.
It is based on dependent type theory, providing a unified framework for
programs, proofs, and mathematical objects. Its main library, \texttt{mathlib},
contains a bevy of formalized theorems spanning broad areas of mathematics.

\section{Case Study}\label{sec:case_study} %
The entire study was conducted within Claude Code using the Sonnet 4.6
model~\cite{claude}. No code was written by a human, and the code was only
minimally inspected. A human performed more extensive evaluation on a
separate server.

The project was started with a minimal description giving the LLM a reference for
the congruence closure algorithm~\cite{closure}.

\begin{quote} \itshape %
Start a project that implements an SMT solver. We need a parser, a SAT solver,
for which we will use the ipasir-up interface hash-consing of formulas. Then
theory specific things, like arbitrary precision library. Setup build by
using cmake. The solver will be written in c++20. Use antlr for parsing (the
grammar is here https://github.com/julianthome/smtlibv2-grammar).
Set up the directory structure also testing. cmake should download the necessary dependencies as needed.

We will start with QF-EUF theory solver implementing the congruence closure as
described in Fast congruence closure and extensions by Robert Nieuwenhuis and
Albert Oliveras (the pdf is in the folder for convenience).
 \end{quote}

This quickly proved too naive.
Despite correctly describing SMT solving at a high level, the LLM did not
implement a correct solver from the start. Most strikingly, the solver did
not handle Boolean connectives at all. This was quickly remedied by
pointing the LLM to the core theory
specification\footnote{\url{https://smt-lib.org/theories-Core.shtml}}. Only
after some time did we notice that \texttt{xor} was not implemented as
n-ary, which required an explicit prompt to fix.

Despite being prompted to use CaDiCaL, the agent instead wrote its own
rudimentary SAT solver. We subsequently gave an explicit prompt to integrate
CaDiCaL via the IPASIR-UP interface~\cite{ipasirup}, which the agent accomplished
without difficulty.

Notably, the agent ran the solver without a timeout, leaving long-running
processes unnoticed in the background. We therefore added to the permanent
instructions of the LLM to always run the solver under the \texttt{timeout}
command.\footnote{Claude Code permits inserting such instructions into
\texttt{CLAUDE.md}.}

\subsection{Bugs and Debugging}

Initial development required numerous bug fixes in the congruence closure
algorithm and parsing. The agent fixed bugs autonomously, once presented with an
input file where the solver gave a wrong answer. Several issues stemmed from the
fact that in SMT2, unlike in first-order logic, there is no distinction between
formulas and terms (for example, $f(\forall x.\,x>0)=3$ is valid in SMT2 as long
as $f$ is from \texttt{Bool} to \texttt{Int}). The following is a commit message
written by the LLM for one of the bug fixes.

\begin{quote}
  \itshape \texttt{link\_bool\_term\_to\_euf}: add mutual-exclusivity clause {-\texttt{eq\_true}, -\texttt{eq\_false}}
       so the SAT solver cannot simultaneously merge \texttt{\_\_bool\_true} and \texttt{\_\_bool\_false},
       which was causing false UNSAT for formulas with Bool-sorted UF arguments.
\end{quote}

The model handles bug fixing well, but---like human developers---benefits from
structured debugging support: fuzzing~\cite{fuzzing} and
delta-debugging~\cite{Zeller2002}. We had the LLM generate a random formula
generator and a differential testing script comparing our solver against a
reference solver, rather than using existing SMT fuzzers~\cite{Brummayer2009}. These
techniques also reduce token usage: without them, the LLM would pinpoint bugs
``manually''---correct but expensive.

\subsection{Preprocessing}\label{sub:preprocessing} %

SMT solvers typically have a preprocessing module for simplifying common
benchmark constructs. We instructed the agent to handle unit propagation as well
as short-circuit simplifications of conjunction and disjunction. Interestingly,
a cursory manual inspection identified that trivial simplifications were
missing, such as that $t=t$ simplifies to true and $\texttt{if } C\texttt{ then
} t\texttt{ else } t$ simplifies to $t$.

An interesting set of benchmarks contained in the SMT-LIB are the equational diamond
problems of the following form.
 \begin{align}
   \label{eq:diamond}&(x_i=z_i\land z_i=x_{i+1})\lor(x_i=v_i\land v_i=x_{i+1}),\ i\in
   1..n\\
   \label{eq:inequality}&x_1\neq x_{n+1}
 \end{align}
The problems are trivially unsatisfiable but difficult for DPLL(T) because it
cannot infer literals not already present in the formula. Here, it would be
useful to have access to the literal $x_i=x_{i+1}$, which trivially follows from
the disjunct~\eqref{eq:diamond} and would quickly give us a contradiction
with~\eqref{eq:inequality}. Not having access to this literal, the solver is
forced to explore $2^n$ combinations, which produce one theory lemma each. One
could explore more advanced techniques to handle such issues
internally~\cite{Barrett2006,Nieuwenhuis2005} but it is also a good candidate
for \emph{preprocessing}. We prompted the agent with a simple prompt.

\begin{quote} \itshape
  Consider SMT problem with disjunctions linking a sequence
  of $x_i$ as $x_i= y_i = x_{i+1}$ or $x_i= z_i = x_{i+1}$. With $x_1\neq
  x_{100}$. This causes issues for lazy SMT and problems of this type are in
  \texttt{QF\_UF/eq\_diamond}. Devise preprocessing technique to overcome the
  exponential behavior on this.
\end{quote}

From this prompt alone, the agent devised a preprocessing technique that
solves the diamond problems instantaneously. For each disjunction, it computes
the EUF-closure of equalities in each branch and extracts those common to all
branches. These common consequences are
added as new unit equality formulas to the formula list. For example, for $(x=y \land
y=z)\lor(x=w \land w=z)$ branch 1 gives the closure: $\{x,y,z\}$, branch 2 gives
the closure: $\{x,w,z\}$ with shared nodes: $\{x,z\}$ and common pair $x=z$.

\subsection{Certification}\label{sub:certification} %
For a satisfiable input formula, the produced model is validated by encoding it
back as an SMT problem and checking it with a reference solver. We asked the
agent to integrate such a check into our fuzzing script. The prompt was bundled
with the specification of the \texttt{get-model} command, for which we used a
part of the SMT2 standard.

\begin{quote}\itshape
  Implement the \texttt{get-model} command, see \texttt{get-model-spec.pdf} for specification create
  a script to test a model, i.e., given an SMT2 file, run our solver, if SAT, add
  the model into the file and call (another) solver to check it's indeed SAT.
\end{quote}

This turned out to be quite easy for the agent. It initially had a bug that it
did not introduce explicit constants to represent the members of the (finite)
universe, but it realized that quickly on its own. Another subtle error
arose during model construction on formulas of the form
\texttt{f(distinct(c0,c2))}---again due to Booleans being treated both as
propositions and as terms.

For unsatisfiable input formulas, we ask the agent to produce a Lean proof. More
advanced and more suitable approaches to proof generation for SMT have been
investigated in the literature~\cite{Barbosa23,BarbosaRKLNNOPV22,AndreottiB26}
but we were mainly interested in whether the agent could formally capture the
reasoning behind its own code.

Since we only rely on QF-EUF, all the relevant primitives readily translate to
Lean primitives, e.g., equality in SMT translates to equality in Lean. We do
not build a low-level Lean proof; instead, we rely on two automation tools:
\begin{itemize}
\item \texttt{grind} to verify theory lemmas -- clauses following directly from congruence closure.
\item \texttt{bv\_decide} to verify the propositional fragment. In order to use \texttt{bv\_decide}, we asked the agent to wrap each proposition into \texttt{decide (...)}
\end{itemize}

Both tools are more powerful than strictly needed, but neither alone can solve
standard SMT problems without extra hints from the solver.

Initially, all types, constants, and SMT assumptions were placed as Lean
hypotheses within a single theorem per SMT problem, with theory lemmas as
\texttt{have} statements. This proved infeasible: some SMT problems have
thousands of assumptions, and Lean struggled to parse even the problem alone.
We instead adopted a less ITP-centric design: SMT types, constants, and
assumptions become Lean \texttt{axioms}; each theory lemma is a separate
\texttt{theorem} (discharged by \texttt{grind}); and the final \texttt{False}
proof collects all of these via \texttt{have} before closing with
\texttt{bv\_decide}.

In particular, for the equational diamond problems (see
\cref{sub:preprocessing}), the agent struggled to grasp the distinction between
theory lemmas and the final proof. It was trying to close propositionally
nontrivial subgoals with \texttt{grind}, which did not work. Eventually, we had
to write down our own example of a Lean proof correctly handling equational
diamonds. Then it was able to incorporate such proofs into its proof export.

\begin{figure}[h]
\begin{lstlisting}[caption={Example of a final version of Lean export with equational diamonds, passed Lean verification. Note that the original statement has only one assert, and the proof only one theory lemma (except for preprocessing), otherwise, we would have multiple axioms, or multiple theorems.}]
import Std.Tactic.BVDecide

noncomputable section
open Classical

axiom U : Type
axiom x0 : U
axiom y0 : U
axiom z0 : U
axiom x1 : U
axiom y1 : U
axiom z1 : U
axiom x2 : U
axiom y2 : U
axiom z2 : U
axiom hyp1 : (((decide (x0 = y0) ∧ decide (y0 = x1)) ∨ (decide (x0 = z0) ∧ decide (x1 = z0))) ∧ ((decide (x1 = y1) ∧ decide (y1 = x2)) ∨ (decide (x1 = z1) ∧ decide (x2 = z1))) ∧ ¬(decide (x0 = x2)))

theorem eq_bridge_0 : decide (x0 = y0) → decide (y0 = x1) → decide (x0 = x1) := by grind
theorem eq_bridge_1 : decide (x0 = z0) → decide (x1 = z0) → decide (x0 = x1) := by grind
theorem eq_bridge_2 : decide (x1 = y1) → decide (y1 = x2) → decide (x1 = x2) := by grind
theorem eq_bridge_3 : decide (x1 = z1) → decide (x2 = z1) → decide (x1 = x2) := by grind
theorem cl_0 : decide (x0 = x1) → decide (x1 = x2) → decide (x0 = x2) := by grind

theorem contradiction : False := by
  have hyp1 := hyp1
  have eq_bridge_0 := eq_bridge_0
  have eq_bridge_1 := eq_bridge_1
  have eq_bridge_2 := eq_bridge_2
  have eq_bridge_3 := eq_bridge_3
  have cl_0 := cl_0
  bv_decide
\end{lstlisting}
\end{figure}

The correctness of the proof mechanism was probably the most challenging task of
this case study. Repeated fuzzing and bug fixing were required for the agent to
correctly implement all steps of the proof. Another obstacle is that the agent
does not grasp the complexity at the Lean proof level. We had trouble
communicating to the agent that only easy steps should be presented to the
\texttt{grind} tactic. At the same time, the agent seems not to be able realize
which steps need to be encoded into the proof. \cref{tab:certified}, in the
upcoming section, presents the experimental results on the whole benchmark set.

\section{Performance}\label{sec:performance} %
We evaluated the developed solver on QF-UF non-incremental benchmarks available
in SMT-LIB\footnote{\url{https://zenodo.org/records/1688772}}. We also
  compare to cvc5~\cite{cvc5} and z3~\cite{z3}. The experiments were performed
  on a machine with two AMD EPYC 7513 32-Core processors and with 514~GiB~RAM
  with 100 problem instances run in parallel with 180~s timeout.

\begin{table}
  \caption{Number of solved instances. A win is counted if the solver is not
  worse than the best solver by 1s.}\label{tab:solved}
  \begin{center}
    \begin{tabular}[c]{lll}
      \textbf{solver}    &\textbf{solved} &\textbf{wins}\\\midrule
      z3                 &\numprint{7500} & \numprint{7491} \\
      cvc5               &\numprint{7494} & \numprint{7308} \\
      llm2smt            &\numprint{7468} & \numprint{6486}\\
      llm2smt-no-prepro  &\numprint{7355} &\numprint{6150} \\
      llm2smt-no-th-prop &\numprint{7478} &\numprint{6593} \\
    \end{tabular}
  \end{center}
\end{table}

\begin{figure}[h]
  \begin{center}
    \includegraphics[width=0.43\textwidth]{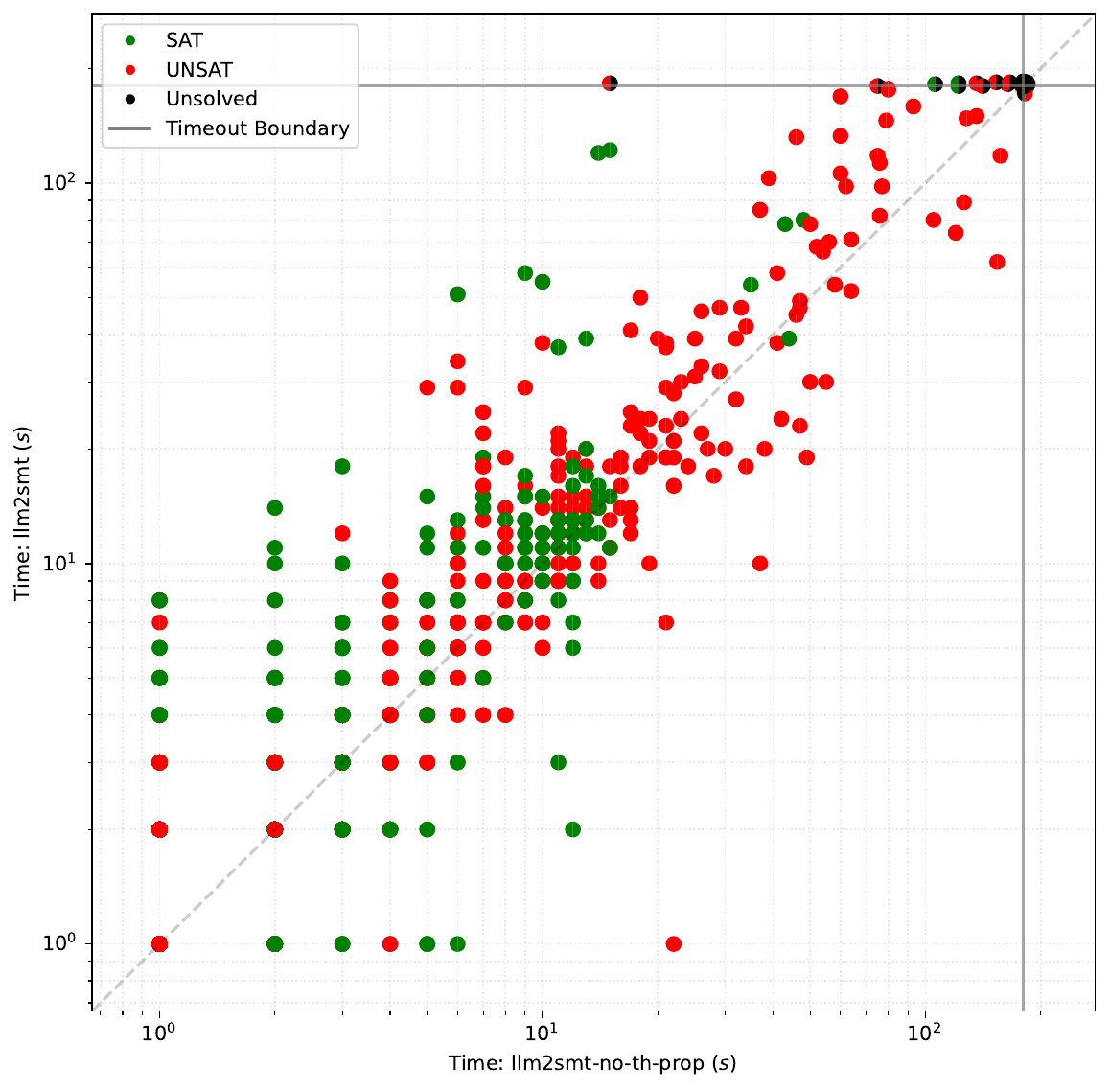}
  \end{center}
  \caption{Comparison with and without theory propagation.}\label{fig:comparison}
\end{figure}

\begin{table}[t]
\centering
\begin{tabular}{lcc}
\toprule
Number of instances & W.\ Preprocessing  & W/o Preprocessing \\
\midrule
Certified           & \numprint{285}     & \numprint{259} \\
Grind Failed        & \numprint{2244}    & \numprint{1949} \\
Timeout             & \numprint{1314}    & \numprint{1472} \\
Maximum Heartbeats  & \numprint{240}     & \numprint{247} \\
Stack Overflow      & \numprint{86}      & \numprint{85} \\
\bottomrule
\end{tabular}
\caption{Overview of the Certification Experiments}\label{tab:certified}
\end{table}

We compare our solver against z3~\cite{z3} and cvc5~\cite{cvc5}, as well as
ablations with preprocessing and theory propagation disabled.
\cref{tab:solved} compares the number of instances solved and gives some
indication of speed---for each solver a win is counted if it is not worse
than the best time by 1s. Interestingly, the version \emph{without} theory
propagation performs slightly better than the one with.
This is likely because propagation introduces overhead without payoff on these
benchmarks.
\cref{fig:comparison} compares the runtime of the version with preprocessing
and the version without, excluding problems solved in under 1 second.
The plot confirms that turning off theory propagation actually makes the solver
faster. However, there are problems where it does make a considerable
difference. This suggests that smarter propagation scheduling could help---for instance,
problems with few function symbols less likely to benefit from it.

\cref{tab:certified} summarizes the certification experiments.
Lean has its own limits on resources. Heartbeats measured the overall time, the
grind tactic is limited on how many times it splits. Additionally, there is a
limit on recursion stack. Further, it is not particularly the appropriate
tool for this task and simply times out on some problems.
This has resulted in a large number of failed certification attempts.
Nevertheless, no erroneous proofs were found.

\section{Observations and Recommendations}\label{sec:observations} %

We conclude with a list of observations and recommendations that hopefully will
be useful to other researchers pursuing a similar goal.

\begin{itemize}
  \item The input language must be very carefully specified (SMT2 in our
    case).
  \item Tools that may be non-terminating or exhibit exponential behavior must
    be run under explicit resource limits.
  \item The agent can debug its own code effectively when given standard tools:
    fuzzing and delta-debugging.
  \item Have the agent commit frequently and produce a regression test for each
    bug fix; it can then use git history to verify and bisect changes.
  \item Despite high overall capability, the agent may fail unexpectedly on
    trivial tasks---e.g., the preprocessor initially did not simplify $s=s$. This phenomenon is known as \emph{jagged
    intelligence}\footnote{\url{https://x.com/karpathy/status/1816531576228053133?s=20}}~\cite{DellAcqua2023}
    and represents a significant challenge for reasoning tools development.
  \item Concrete examples and tests work best to improve the agent's
    performance and reliability.
  \item Proof-generation capabilities appeared to be rather hard to achieve for 
    the agent.
\end{itemize}

\section{Conclusion and Future Work}\label{sec:future} %
The central question we posed was: \textit{Can LLMs develop automated reasoning
tools?} The answer is a qualified yes. With appropriate scaffolding---systematic
fuzzing, delta-debugging, explicit resource limits, and careful specification of
input formats---a coding agent can produce a competitive SMT solver from scratch,
solving nearly as many benchmarks as mature solvers such as z3 and
cvc5---admittedly on a single theory. The agent is capable of implementing
techniques described in the research literature and debugging its own code when
provided with concrete failing examples.

Directing the agent toward established techniques such as delta-debugging and
fuzzing proved beneficial. Proof generation was the most challenging aspect: the
agent must understand not only the solver logic but also the expectations of the
proof checker (Lean in our case), which required significant human guidance and
still fails on a number of problems. The agent does grasp the structure of proof
construction---in several cases it successfully corrected its own invalid
proofs---yet it struggles to faithfully encode all computational information
produced during solving into the final proof certificate.

In future work we plan to extend the solver with additional theories and to
incorporate more systematic human code review into the agent's feedback loop,
with the goal of identifying and guiding improvements to the generated
implementation (human-in-the-loop).

\end{document}